# Quantum Octets in Air Stable High Mobility Two-Dimensional PdSe$_2$

Yuxin Zhang[1], Haidong Tian[1], Huaixuan Li[2,3], Chiho Yoon[2], Ryan A. Nelson[4], Ziling Li[1], Kenji Watanabe[5], Takashi Taniguchi[6], Dmitry Smirnov[7], Roland Kawakami[1], Joshua E. Goldberger[4], Fan Zhang[2], Chun Ning Lau[1*]

[1] Department of Physics, The Ohio State University, Columbus, OH 43210.
[2] Department of Physics, The University of Texas at Dallas, 800 West Campbell Road, Richardson, Texas 75080-3021, USA
[3] Department of Physics, Carnegie Mellon University, Pittsburgh, Pennsylvania 15213, USA
[4] Department of Chemistry and Biochemistry, The Ohio State University, Columbus, OH 43210.
[5] Research Center for Electronic and Optical Materials, National Institute for Materials Science, 1-1 Namiki, Tsukuba 305-0044, Japan
[6] Research Center for Materials Nanoarchitectonics, National Institute for Materials Science, 1-1 Namiki, Tsukuba 305-0044, Japan
[7] National High Magnetic Field Laboratory, Tallahassee, FL 32310

**Two-dimensional (2D) materials have drawn immense interests in scientific and technological communities, owing to their extraordinary properties that are profoundly altered from their bulk counterparts and their enriched tunability by gating, proximity, strain and external fields. For digital applications, an ideal 2D material would have high mobility, air stability, sizable band gap, and be compatible with large scale synthesis. Here we demonstrate air stable field effect transistors using atomically thin few-layer PdSe$_2$ sheets that are sandwiched between hexagonal BN (hBN), with record high saturation current >350 µA/µm, and field effect mobilities 700 and 10,000 cm$^2$/Vs at 300K and 2K, respectively. At low temperatures, magnetotransport studies reveal unique octets in quantum oscillations, arising from 2-fold spin and 4-fold valley degeneracies, which can be broken by in-plane and out-of-plane magnetic fields toward quantum Hall spin and orbital ferromagnetism.**

Recently, a class of 2D materials with puckered pentagonal lattice structures[1-4] has arrived at scene. These materials have broken sublattice symmetry, and have been predicted to host exotic new properties partly due to the in-plane and cross-plane anisotropy[5-7], including much larger band gap variation, axis-dependent conduction polarity, enhanced spin−orbit coupling[8], and nonsymmorphic symmetry-enforced band topology in the monolayer limit[9]. One such example is PdSe$_2$ [10-17]. Notably, bulk PdSe$_2$ are reported to display air stability[18], a widely tunable band gap that varies from 0.5 eV in bulk to 1.3eV for monolayers[1, 18-20], ambipolar transport[21], superconductivity upon transforming to a cubic polymorph under high pressure[22], and superior optical and thermoelectric properties[23-25]. Wafer-scale synthesis of few-layer PdSe$_2$ has already been developed[17, 26, 27]. Mobilities of up to 200 cm$^2$/Vs has been reported, albeit only under large source-drain bias of 1-2V [1, 18, 28]. Thus, as a recent addition to the family of 2D materials, PdSe$_2$ holds great promise for digital electronic, thermoelectric, and optoelectronic applications.

---

* Email: lau.232@osu.edu

Despite the increasing interest in NMDCs in the past few years, transport measurements so far are performed only under high bias and at room temperature, with little optimization with regard to contact resistance or systematic investigation of device behavior. Here we report transport studies of atomically thin PdSe$_2$ field effect transistors that are stable in ambient conditions for > 1 month, record high saturation current among atomically thin semiconductors, as well as record high field effect mobility. This ultra-high mobility enabled the first experimental observation of Shubnikov-de Hass oscillation and the quantum hall effect in this pentagonal 2D material. Interestingly, the Landau fan reveals an 8-fold degeneracy, arising from 2-fold spin and 4-fold valley degeneracies[29]; increasing in-plane ($B_\parallel$) and out-of plane ($B_\perp$) magnetic fields leads to Landau level crossings and broken spin and valley symmetries. Those observations indicate a unique band structure and spin-valley interplay in the pentagonal PdSe$_2$.

Each unit cell of PdSe$_2$ consists of two inversion-symmetric planes of d$^8$ Pd$^{2+}$ ions bonded in a square-planar-like coordination with diselenide dianions (Fig. 1a-b), yet the two adjacent layers are related by glide-mirror symmetry. The lattice constants are $a$=5.74Å, $b$=5.86Å, and $c$=7.69 Å as measured with x-ray diffraction (see Supplementary Information (SI)). Note that the unit cell of the bulk consists of two PdSe$_2$ layers separated by a van der Waals gap; from atomic force microscope measurements, each monolayer step has a height of 5.2Å (see SI). When projected onto a plane, each layer consists of a puckered network of pentagonal rings. First-principles calculations show that PdSe$_2$ few-layers are indirect-gap semiconductors, with a 1.06 eV band gap for monolayer and a <0.1 eV band gap for heptalayer (Fig. 1c-d, and see SI). Their conduction band minima are located close to the $M$ point in the Brillouin zone and thus form four symmetry-related valleys. This valley quartet are universal in our few-layer calculations, although the band gap decreases with increasing the thickness as approaching to the bulk limit (see SI). (Note that the GW method yields larger band gaps for quasiparticle bands [29].)

Bulk PdSe$_2$ crystals were purchased commercially, or grown via vertical Bridgman in Se flux (see SI), and exfoliated into atomically thin layers onto insulating substrates. Two types of devices are fabricated. We first fabricate "bare" devices on Si/SiO$_2$ substrates, by transferring few-layer graphene on few-layer PdSe$_2$ sheets as electrodes, and depositing Cr/Au contacts on graphene. The red curve in Fig. 1e displays the four-terminal resistance $R_{xx}$ as a function of back gate voltage $V_{bg}$ for an as-fabricated 4 nm-thick device. The device appears to be ambipolar, with maximum $R_{xx}$~600 kΩ located at $V_{bg}$=7.8V. In the linear response regime, its electron and hole field effect mobilities $\mu_{FE}$=(1/e)(d$\sigma$/dn) are ~24 and 11 cm$^2$/Vs, respectively (here $\sigma$ is the 2D conductivity, $n$ the charge density, and $e$ the electron charge). To test the device's stability in ambient condition, we leave the device in air and monitor the transfer curve as a function of time. Surprisingly, despite previous claims of PdSe$_2$'s air stability[1, 20, 30], the device deteriorates steadily with time – $\mu_{FE}$ decreases, while the resistance maximum shifts to the right; this $p$-doping aging effect likely arises from oxidation into PdSe$_2$O$_x$. By day 27, the device loses all response to gate, and $R_{xx}$=350 kΩ. We find that thermal annealing at 200°C in vacuum restores the mobility of the electron-doped regime, though the resistance maximum remains at $V_{bg}$<-40V, indicating that the device is electron-doped, which could arise from the formation of Se vacancies.

To fabricate air stable, high performance field effect transistor devices, we take advantage of hexagonal BN (hBN) layers, which have been shown to provide protection for air-sensitive materials such as phosphorene[31], InSe[32] and $CrI_3$[33]. $PdSe_2$ sheets that are 3-8 layers thick are contacted by few-layer graphene leads and sandwiched between hBN layers (Fig. 1b). The global $Si/SiO_2$ back gate tunes both the charge density $n$ of $PdSe_2$ layer and the contact between graphene and $PdSe_2$, while the top gate covers only the channel region and tunes only $n$ therein. As-fabricated hBN-encapsulated devices have high resistance, >~ MΩ, for $V_{bg}$≤0; as $V_{bg}$ increases from 0, four-terminal resistance drops rapidly to a few kΩ upon electron doping. These devices are very stable in air, with no degradation in mobility and only slight hole doping after 29 days (Fig. 1f).

We now focus exclusively on hBN-sandwiched $PdSe_2$ devices. Fig. 1g-h displays the two-terminal current-voltage (I-V) characteristics of a $PdSe_2$ device that is ~2.5 nm thick or ~5 layers at room temperature. When $V_{bg}$<0V, the device is intrinsic and the zero-bias resistance is ~350 MΩ and saturation current $I_{sat}$ is a few nA/μm. With increasing doping, the two-terminal resistance $R_{2T}$ decreases and $I_{sat}$ increases. At $V_{bg}$=70V and $V_{tg}$=10V, $I_{sat}$ ~350 μA/μm, which is among the highest reported for atomically thin 2D semiconductors. We also note that here the contact resistance is still quite prominent (e.g. when highly doped, $R_{2T}$ and $R_{xx}$ are ~35 kΩ and 7 kΩ, respectively), thus we expect that $I_{sat}$ can reach > mA/ μm by minimizing contact resistance of the future generation of devices.

To examine the dominant scattering mechanism, we plot the average field effect mobility between 4.5 and 6.5x $10^{12}$ $cm^{-2}$ as a function of temperature $T$ for two separate devices. As shown by Fig. 2a, the room temperature $\mu_{FE}$ is ~150 and 780 $cm^2$/Vs for the 5-layer and 7-layer devices, respectively. At high temperatures, $\mu_{FE}$ increases with decreasing temperature, with a power-law behavior $T^{-\alpha}$. Fitting the data yields $\alpha$ =1.3 and 1.6 for ~5-layer and 7-layer device, respectively, indicating that the main scattering mechanism is phonon scattering. For $T$<30K, $\mu_{FE}$ saturates to 2,000 and 10,000 $cm^2$/Vs, respectively. These mobility values, which are 1-3 orders of magnitude higher than prior reports[1,19,23][34], suggests that the devices have reached the regime where the mobility bottleneck is scattering by intrinsic defects and/or impurities.

We now turn to the magneto-transport measurements at the cryogenic temperatures. Here we focus on a sample that is ~3.5 nm thick or ~7 layers. Fig. 2b shows the background subtracted longitudinal resistance $\Delta R_{xx}$ plotted versus $V_{tg}$ and the perpendicular magnetic field $B_\perp$, while the back gate voltage is maintained at $V_{bg}$ =65V. Clear Shubnikov-de Hass (SdH) oscillations from $PdSe_2$ start to be resolved around 4T, indicating quantum mobility of 2500 $cm^2$/Vs. Interestingly, the Landau fan features resistance minima at filling factors $\nu=nh/eB_\perp$ that are integer multiples of 8 (Fig. 2b-c); such quantum Hall octets have not been observed in other 2D materials. To account for this 8-fold degeneracy, we note hat the conduction band bottom occurs near the $M$ point in the rectangular Brillouin zone, and that the glide and screw symmetries dictate a 4-fold valley degeneracy (Fig. 2c inset). Additionally, near the conduction band bottom, the spin-orbit coupling is negligibly small, as can be seen in Fig. 1c-d and Fig. S3-4, the bands have a 2-fold spin degeneracy even for even-layer systems in which inversion symmetry is broken. Consequently, the charge carriers in PdSe$_2$ are endowed with the observed 8-fold degeneracy.

The effective mass of the charge carriers is extracted by evaluating the amplitude of the SdH oscillations with temperature. Fig. 2d plots $R(V_{tg})$ at $T$ ranging from 1.7K to 7K at $B_\perp$=7.5T, and Fig. 2c the oscillation amplitude as a function of temperature. Fitting the data to the Lifshitz-Kosevich formula $\Delta R \propto \frac{\chi}{\sinh(\chi)}$, where $\chi = \frac{2\pi^2 k T m^*}{\hbar e B_\perp}$, $k$ is the Boltzman constant, $\hbar$ is the reduced Planck constant, and $T$ is the temperature, we obtain an effective mass $m^*$=0.29$m_e$, here $m_e$ is the bare electron mass in vacuum. This result is in excellent agreement with previous theoretical calculations[29].

At higher fields, degeneracy of the quantum Hall octets is partially lifted. Major resistance valleys occur at filling factors $\nu$=8$N$, where $N$=1, 2, 3… is an integer denoting the Landau level (LL) index; between the major valleys, minor resistance dips are visible at $\nu$=4$N_{odd}$, where $N_{odd}$= 1, 3, 5… is an odd integer. To gain insight into the nature of this quantum Hall ferromagnetism, we perform measurements in tilted magnetic fields, since the valley (orbital) degrees of freedom depends only on the perpendicular component of $B$, whereas the spin couples to the total magnetic field $B_t$. Fig. 3a-c plots the $R(B_t, \nu)$ for data acquired at angles $\theta$=0, 35.6° and 48.2°, respectively, and line traces of $R(V_{tg})$ are shown in Fig. 3d. At $\theta$=0 (i.e. $B_t$=$B_\perp$ and $B_\parallel$=0), the QH states at $\nu$=8, 16, 24 and 32 are resolved at lower fields, and those at $\nu$=4, 12, 20 and 28 are resolved at higher fields with shallower minima, indicating that the cyclotron gaps for the former (latter) are larger (smaller). At $\theta$=35.6°, the adjacent states are approximately equally resolved, suggesting that the gaps at $\nu$=8$N$ ($\nu$=4$N_{odd}$) decrease (increase) with increasing $B_\parallel$. Upon further increase of $B_\parallel$, at $\theta$=48.2°, only the QH states at $\nu$=4$N_{odd}$=12, 20, 28 and 36 are resolved, *i.e.* the cyclotron and Zeeman gaps become equal in magnitude, and the LLs with indices ($N$, ↑) and ($N$+1, ↓) cross each other. This is the so-called first coincidence angle, at which LLs with opposite spin indices cross.

These data enable us to determine the schematic of LL resolution. At low field, the 8-fold degenerate LLs are separated by cyclotron gaps $E_c$=$\hbar\omega_c(N+\frac{1}{2})$, where $\omega_c=\frac{eB_\perp}{m^*}$ is the cyclotron frequency. Increasing $B_\perp$ first lifts the spin degeneracy, giving rise to Zeeman-split LLs with energy difference $E_Z$=$g\mu_B B_t$, where $\mu_B$ is the Bohr magneton and $g$ the effective electron $g$-factor including the Coulomb interaction correction. In titled field measurements, increasing $B_\parallel$ further enhances the Zeeman splitting, giving rise to LL crossing at large $B_\parallel$ (Fig. 3e). Using this schematic, we can extract the effective $g$-factor from the coincidence angle at which the cyclotron and Zeeman gaps are equal in magnitude ($\theta$=48.2°). This condition yields $g = \frac{\hbar e}{\mu_B m^*}\frac{B_\perp}{B_t} = \frac{2\cos(\theta_2)}{m^*/m_e}$=4.60, respectively, using $m^*$=0.29$m_e$ obtained earlier. This enhancement from the bare value of $g$=2 in a nearly spin-orbit-coupling-free system substantiates the presence of electron-electron interaction effect and the resulting quantum Hall spin ferromagnetism.

Finally, we perform measurements at even higher $B_\perp$ fields up to 29T and observe that the 4-fold valley degeneracy is broken (Fig 4a). This quantum Hall orbital ferromagnetism is clearly illustrated by the emergence of additional resistance minima between filling factor 4, 8 and 12 (Fig 4b). Future study is required to show whether these quantum Hall states are valley coherent phases breaking translational symmetry or valley polarized phases exhibiting in-plane ferroelectricity[35]. Notably, quantum hall effects down to the lowest LL is resolved. attesting to the excellent quality of our sample. The entire sequence of degeneracy breaking is schematically illustrated in Fig. 4c.

In conclusion, we demonstrate that few-layer pentagonal PdSe$_2$, when sandwiched between hBN layers, is an excellent 2D semiconductor, boasting air stability, record high saturation current, exceedingly high field effect mobility, and quantum Hall octets and ferromagnetism in magnetic fields. Our work paves the way for future electronic, photonic, and topological applications of this pentagonal material.

**Acknowledgement**

We thank Dima Shcherbakov for helpful discussion. This work is supported by NSF/DMR 2128945. H.T. is supported by DOE BES Division under grant no. DE-SC0020187. R.A.N. and J.E.G. acknowledge the Air Force Office of Scientific Research for funding from grant number FA9550-21-1-02684. Z.L. and R.K.K. acknowledge support from AFOSR/MURI project 2DMagic (FA9550-19-1-0390) and the US Department of Energy (DE-SC0016379). A portion of this work was performed at the National High Magnetic Field Laboratory, which is supported by the National Science Foundation through NSF/DMR-1644779 and the State of Florida. K.W. and T.T. acknowledge support from the JSPS KAKENHI (Grant Numbers 20H00354, 21H05233 and 23H02052) and World Premier International Research Center Initiative (WPI), MEXT, Japan.

Figure 1. Atomic and band structures, and room temperature transport data of few-layer $PdSe_2$ field effect transistors. (a). Top and side view of the lattice structure of 2D puckered $PdSe_2$. Blue and yellow spheres represent the Pd and Se atoms, respectively. (b). The device schematics (top) and an optical image of the stack (below). (c-d). Band structures of 7-layer $PdSe_2$ without and with spin orbit coupling included in the calculation, respectively. (e-f). Transfer characteristics $R_{xx}(V_{bg})$ of $PdSe_2$ device on $SiO_2$ and encapsulated within hBN, respectively, after periods of time in air. Inset in f: optical image of ahBN/PdSe2/hBN stack. (g-h). I-V characteristics of a 6-layer device at different gate voltages, showing saturation current of ~350 µA/µm when highly doped.

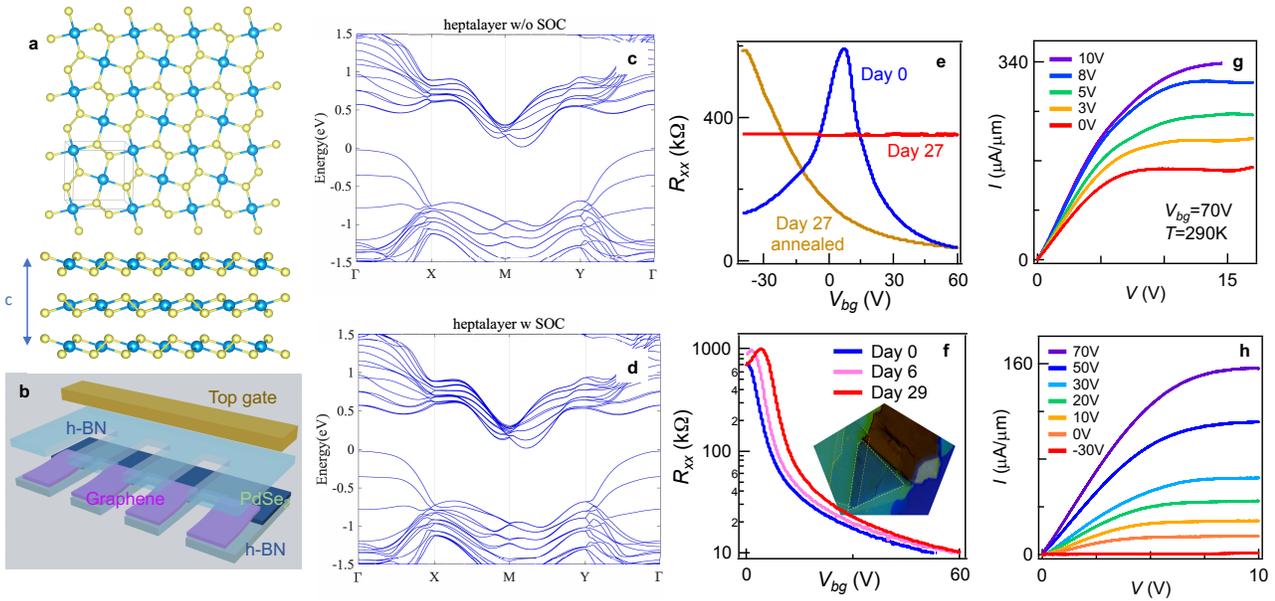

Figure 2. Transport data of PdSe$_2$ transistors at low temperatures. (a). Field effect mobility as a function of temperature of two PdSe$_2$ transistors that are ~3.5 nm(green) and 2.5 nm (red) in thickness. Dashed lines are fits to $T^{-\gamma}$. (b). SdH oscillations of a 7-layer PdSe$_2$ devices vs $V_{tg}$ and $B_\perp$. Inset: Energy contours of conduction band of 7-layer PdSe$_2$ in the $k_x$-$k_y$ plane, showing the four valleys. To better showcase the valleys, segments of the axes are omitted (indicated by dotted lines). (c). $R(V_{tg})$ at $B_\perp$=7.5T and different temperatures. (d). Normalized amplitude of oscillations vs temperature, fitted to the Lifshitz-Kosevich equation.

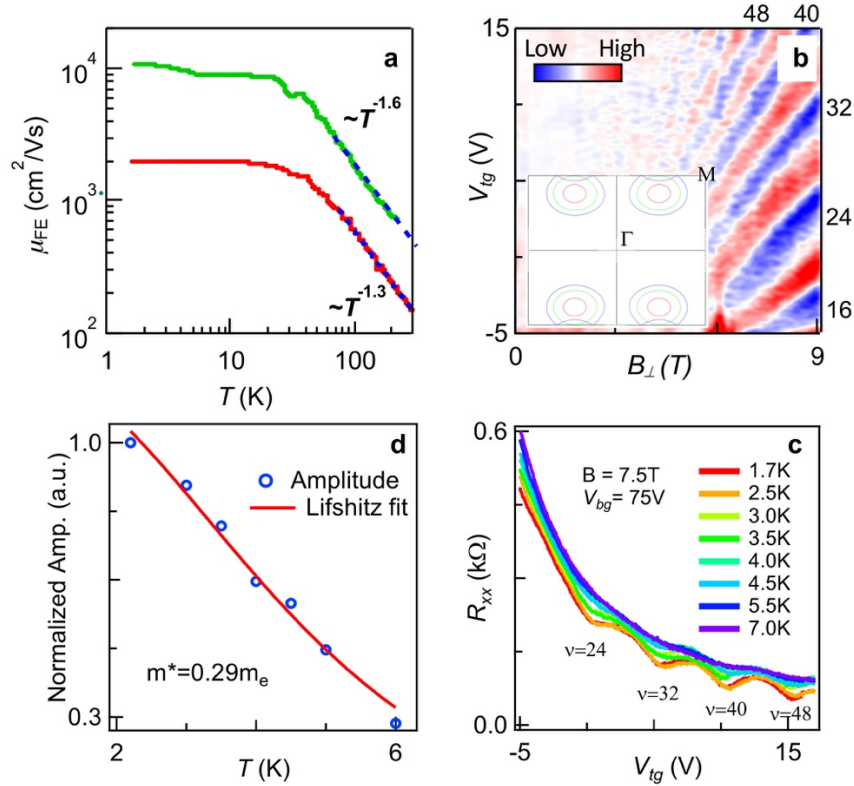

Figure 3. Magnetotransport data in tilted magnetic fields for the 7-layer device. (a-c). $R$ vs total magnetic field and filling factor $\nu$ at different tilting angles. (d). Line traces of $R(V_{tg})$ at $B_\perp=12T$ and different tilting angles. Black numbers indicate filling factors. (e). Schematics of the evolution of Landau levels with in-plane magnetic field while $B_\perp$ is kept constant.

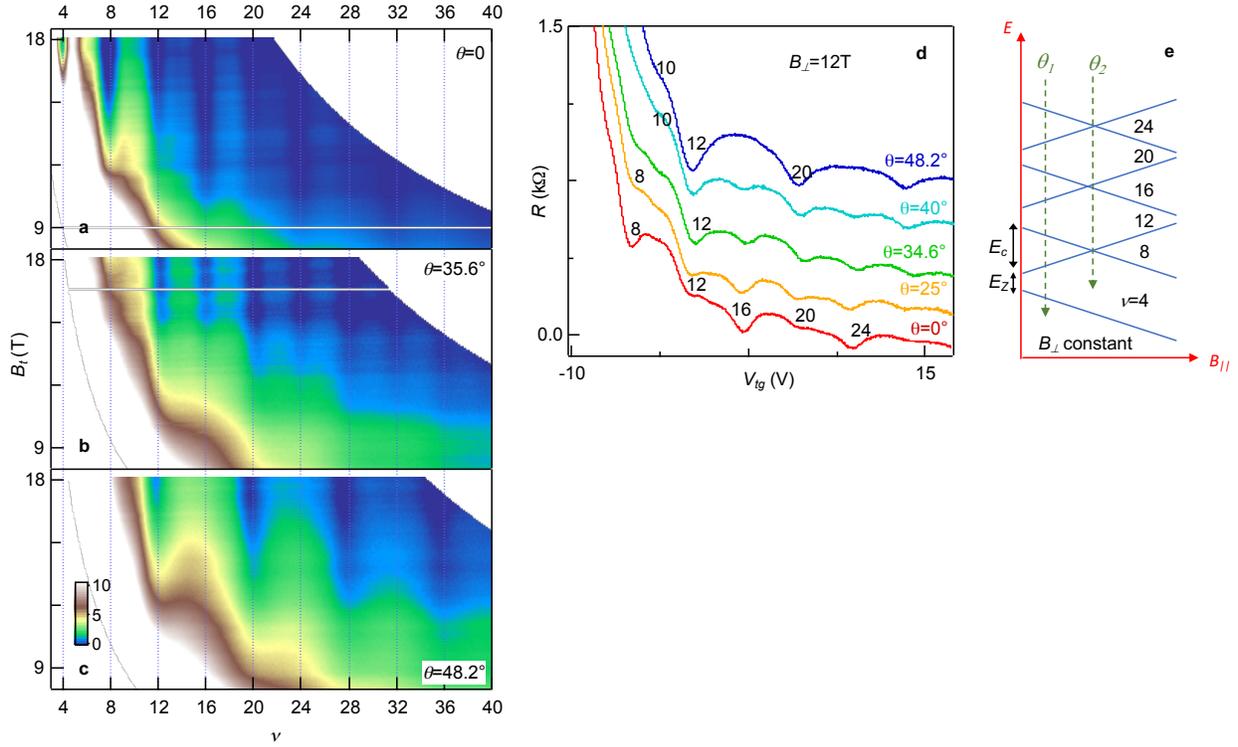

Figure 4. Magnetotransport data at high magnetic field. (a). $R_{xx}(V_{tg}, B_\perp)$ at $V_{bg}=75$V from $B_\perp=8$ to 29T. (b). Line traces $R_{xx}(V_{tg})$ at $B_\perp=18$ and 26T, respectively. Filling factors are labeled on the corresponding resistance dips. (c). Schematics of the resolution sequence of the cyclotron, Zeeman and valley gaps.

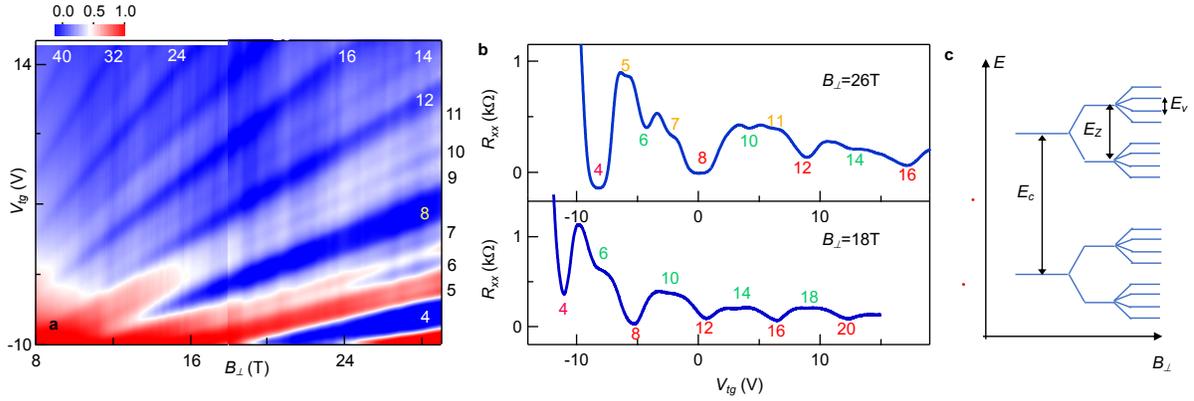